\documentclass[aps,prc,showpacs,nofootinbib]{revtex4}
\usepackage{epsfig}
\usepackage{graphicx,amsmath}
\newcommand\ba{\begin{eqnarray}}
\newcommand\ea{\end{eqnarray}}

\newcommand{\be}{\begin{equation}}
\newcommand{\ee}{\end{equation}}
\newcommand{\PANDA}{$\overline{\rm P}$ANDA }
\begin{document}
\title{Compared analysis of proton electromagnetic form factors in space-like and time-like regions}
 
\author{E. Tomasi-Gustafsson}
\affiliation{IRFU,SPhN, 91191 Gif-sur-Yvette and CNRS/IN2P3, IPNO, UMR 8608, 91405 Orsay, France }


\begin{abstract}
New possibilities of high precision measurements of hadron form factors in annihilation and scattering reactions over an unexplored kinematical region  suggest a compared analysis, in view of a global description of the nucleon structure.
\end{abstract}
\maketitle
\section{Introduction}
 
The study of hadron electromagnetic form factors (FFs) is a very active field of high energy physics since a few decades, as FFs are fundamental quantities which contain information on the internal structure of composite particles. They constitute a privileged playground for the test of theoretical models, and should reflect the transition from the non perturbative regime, where effective degrees of freedom describe the nucleon structure, to the asymptotic region  where QCD applies. The possibility to transfer high momenta, and therefore to access small internal distances, allows to test pQCD predictions, such as quark counting rules and helicity conservation. 

In the space-like region, high precision measurements in an extended kinematical range are an important part of the present and future experimental program at Jefferson Laboratory (USA). In the time-like region, a program is foreseen by the \PANDA collaboration at FAIR (Germany), using high intensity antiproton beams up to 15 GeV kinetic energy. Similar studies are also discussed as part of the experimental program at electron positron colliders, in Frascati (Italy), Novosibirsk (Russia), Beijing (China).

The traditional way to measure electromagnetic hadron FFs is based on elastic electron proton scattering $e^-+p\leftrightarrow e^-+p$ and on the annihilation reactions related by crossing symmetry $e^++e^-\leftrightarrow p+\bar p$, assuming that the interaction occurs through the exchange of one virtual photon, of mass $q^2$. In recent years, very surprising results have been obtained in $ep$ elastic scattering, due to the possibility of applying the polarization method \cite{Re68}: the electric and magnetic distributions inside a proton do not have the same dipole dependence, as a function of $q^2$ \cite{Jo00}, as it was previously assumed.

The understanding and the interpretation of the data at large momenta in the full kinematical region requires to investigate carefully not only the nucleon structure but also the reaction mechanism. The simple extrapolation of models and methods should be taken very carefully. A large debate recently arose, due to inconsistencies among form factors extracted from polarized and unpolarized experiments in space-like region (for a review, see \cite{Pe07}). As no bias has been found in both types of experiment and as the extraction of form factors follows the same formalism (based on one-photon exchange), possible explanations are related to higher order radiative corrections.

The measured observables are the differential cross section in unpolarized $ep$ scattering and the ratio of the longitudinal to transverse proton polarization in elastic scattering of longitudinally polarized electron on an unpolarized proton target. Radiative corrections are very large for the unpolarized cross section, and are neglected in polarization experiments. High order corrections have not yet been applied to the data. The presence of two-photon exchange would induce a more complicated spin structure of the matrix element and it has been discussed in the frame of a compared analysis of space-like \cite{ETG,Re04} and time-like data \cite{Ga06,Ga05,ETG05}. 

\section{Formalism }
\subsection{Space -like region} 
Assuming one-photon exchange  the reduced elastic differential cross section for $ep$ elastic scattering, may be written as \cite{Ro50}:
 \ba
\sigma_{red}(\theta,Q^2)&=&\epsilon(1+\tau)\left [1+2\displaystyle\frac{E}{m}\sin^2(\theta/2)\right ]\displaystyle\frac
{4 E^2\sin^4(\theta/2)}{\alpha^2\cos^2(\theta/2)}\displaystyle\frac{d\sigma}{d\Omega}=\tau G_M^2(Q^2)+\epsilon G_E^2(Q^2),\\ 
\epsilon&=&[1+2(1+\tau)\tan^2(\theta/2)]^{-1},~\tau=Q^2/(4m^2),~Q^2=-q^2
\label{eq:sigma}
\ea
where $\alpha=1/137$, $m$ is the proton mass, $E$ and $\theta$ are the incident electron energy and the scattering angle of the outgoing electron in the laboratory system, respectively. $G_M(Q^2)$ and $G_E(Q^2)$ are the magnetic and the electric proton FFs, functions of $Q^2$, only. Measurements of the elastic differential cross section at different angles for a fixed value of $Q^2$ allow $G_E(Q^2)$ and $G_M(Q^2)$ to be determined as the slope and the intercept, respectively, from the linear $\epsilon$ dependence (\ref{eq:sigma}). The normalization is chosen in order to have static values proportional to the proton electric charge and magnetic moment $\mu$, respectively $G_E(0)=1$ and $G_M(0)=\mu$. 

From unpolarized cross section measurements the determination of $G_E$ and $G_M$ has been done up to $Q^2\simeq 8.8 $ GeV$^2$ \cite{An94} and $G_M(Q^2)$ has been extracted up to $Q^2\simeq 31$ GeV$^2$ \cite{Ar86} under the assumption that $G_E=0$, and it is often approximated, for practical purposes, according to a dipole form: $G_D(Q^2)=  (1+Q^2/0.71\mbox{~GeV}^2)^{-2}$. Polarization transfer measurements suggest a monotonical decrease of the ratio $\mu G_E(Q^2)/G_M(Q^2)$ with $Q^2$:
\be  
\mu G_E/G_M=1 \mbox{~for~} Q^2<0.4~[(\mbox{GeV/c})^2],~
\mu G_E/G_M= 1.0587 -0.14265\mbox{~for~} Q^2\le 6~[(\mbox{GeV/c})^2],
\label{eq:polar}
\ee
at larger $Q^2$, at least up to $Q^2\sim 6$ GeV$^2$, deviating from unity as $Q^2$ increases \cite{Jo00}.

\begin{figure}
\begin{center}
\includegraphics[width=10cm]{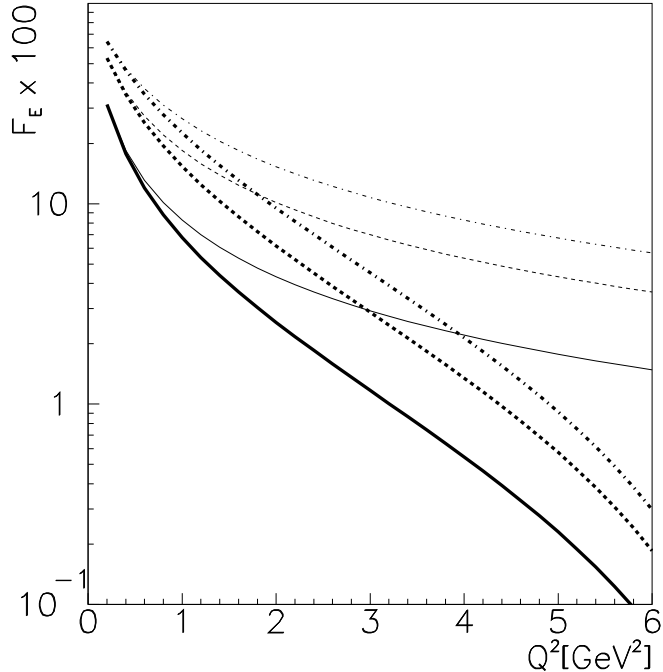}
\caption{\label{Fig:fig2} Contribution of the $G_E(Q^2)$ dependent term to the reduced cross section (in percent) for $\epsilon=0.2$ (solid line), $\epsilon=0.5$ (dashed line), $\epsilon=0.8$ (dash-dotted line), in the hypothesis of FF scaling (thin lines) or following Eq. (\protect\ref{eq:polar}) (thick lines).}
\end{center}
\end{figure}

At large $Q^2$ the contribution of the electric term to the cross section becomes very small, as the magnetic part is amplified by the kinematical factor $\tau $. This is illustrated in Fig. \ref{Fig:fig2}, where the ratio of the electric part to the reduced cross section  $F_E=\epsilon G_E^2(Q^2)/\sigma_{red}(\theta,Q^2)$, is shown as a function of $Q^2$, for different values of $\epsilon$. The different curves correspond to different values of $\epsilon$, assuming FFs scaling (thin lines) or in the hypothesis of the linear dependence of Eq. (\ref{eq:polar}) (thick lines). In the second case, one can see that, for example, for $\epsilon=0.2$ the electric contribution becomes lower than $3$\% starting from 2 GeV$^2$. This number should be compared with the absolute uncertainty of the cross section measurement. When this contribution is larger or is of the same order, the sensitivity of the measurement to the electric term is lost and the extraction of $G_E(Q^2)$ becomes meaningless.

Higher precision can be obtained in polarization experiments. As it was firstly shown in \cite{Re68}, measuring the polarization of the outgoing proton in the scattering of longitudinally polarized electrons on an unpolarized proton target, gives access to an interference term which contains the product $G_E(Q^2)G_M(Q^2)$ and it is more sensitive to a small electric contribution than the cross section itself. The following expressions hold for the transverse and longitudinal components 
$P_t$ and $P_{\ell}$ of the proton polarization vector (in the scattering plane) in terms of the proton electromagnetic FFs:
\ba
DP_t&=&-2\lambda \cot \displaystyle\frac{\theta}{2} \sqrt{\displaystyle\frac{\tau}{1+\tau }}G_EG_M,
DP_{\ell}=\lambda\displaystyle\frac{E+E'}{m}\sqrt{\displaystyle\frac{\tau}{1+\tau }}G_M^2,~\\
D&=&2\tau G_M^2+\cot^2 \displaystyle\frac{\theta}{2} \displaystyle\frac{G_E^2+\tau G_M^2}{1+\tau }.
\label{eq:fi}
\ea
where $E'$ is the scattered electron energy and $D$ is proportional to the differential cross section with unpolarized particles. So, for the ratio of these components one can find the following formula:
\begin{equation}
\displaystyle\frac{P_t}{P_\ell}= - 2\cot \displaystyle\frac{\theta}{2} \displaystyle\frac{m}{E+E'}\displaystyle\frac{G_E(q^2)}{G_M(q^2)}
\label{eq:final}
\end{equation}
which shows the direct link between the polarization components of the recoil proton and the electric and magnetic FFs. 

The results obtained with such technique display a large precision compared to the Rosenbluth separation, due to the large sensitivity to the electric FF. Moreover, the electron beam polarization as well as the analyzing powers of the polarimeter cancel in the ratio, reducing the systematic errors. 

\subsection{Time-like region}
Due to unitarity, in TL region hadron FFs are complex functions of $q^2$. The unpolarized cross section depends on their moduli, the measurement of which is, in principle, simpler than in SL region where the Rosenbluth separation requires at least two measurements at fixed $q^2$ and different angles implying a change of incident energy and scattered electron angle at each $q^2$ point. In TL region, the individual determination of $|G_E|$ and $|G_M|$ requires the measurement of the angular distribution of the outgoing leptons, at fixed total energy $s=q^2$. All the information, of the nucleon structure as well as of the reaction mechanism, as discussed below, is contained in the differential cross section.
%

The differential cross section for the annihilation process
\be 
\bar p + p\to\ell^++ \ell^-,~  \ell=e,\mu 
\label{eq:eq1}
\ee 
was first obtained in ref. \cite{Zi62}. It is expressed in terms of the proton electromagnetic FFs as:
\be
\displaystyle\frac{d\sigma}{d(cos\theta)} = \displaystyle\frac{\pi\alpha^2}{8m^2\tau\sqrt{\tau(\tau-1)}}
\left [ \tau |G_M|^2(1+\cos^2\theta)+ |G_E|^2\sin^2\theta
\right ], 
\label{eq:eq2}
\ee
where $\theta$ is the electron production angle in the center of mass system (CM). The $\cos^2\theta$ dependence of Eq. (\ref{eq:eq2}) results directly from the one-photon exchange mechanism, assuming that the spin of the photon is equal to one and that the electromagnetic hadron interaction satisfies $C$-invariance. This corresponds, by crossing symmetry, to the linear Rosenbluth $\cot^2(\theta/2)$ dependence \cite{Re99}.

\begin{figure}
\begin{center}
\includegraphics[width=14cm]{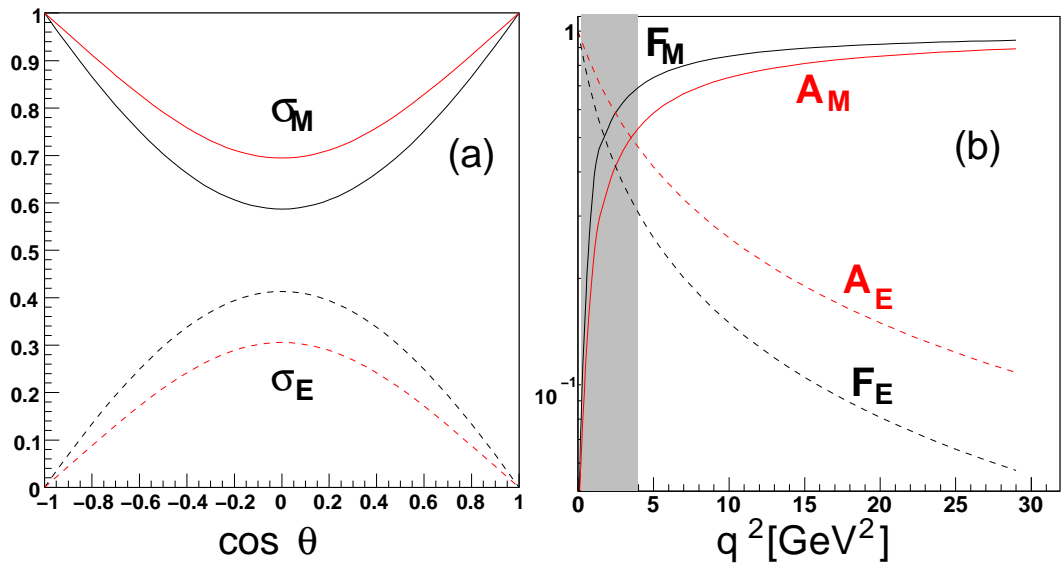}
\vspace*{.2 truecm}
\caption{(a) Relative contribution of the electric $\sigma_E$ (dashed lines) and magnetic $\sigma_M$ (solid lines) terms to the differential cross section for $\bar p+p\to e^++e^-$, as functions of $\cos\theta$ for two different values of $q^2$: 5 GeV$^2$ (red lines) and  $q^2$ =8 GeV$^2$ (black lines);
(b) Relative contribution of the electric (dashed lines) and magnetic (solid lines) terms to the total cross section $F_E$ and $F_M$ (black lines) and to the angular asymmetry, $A_E$  and $A_M$ (red lines).}
 \label{Fig:ang}
 \end{center}
\end{figure}

The electric term is accompanied by a dependence in $\sin^2\theta$. It means that, whatever is the model used for $|G_E(q^2)|^2$, it has a maximum at $\cos\theta=0$ and
vanishes at $\cos\theta=\pm 1$. 
The magnetic term has a maximum at $\cos\theta=\pm 1$, which equals to $2\tau |G_M(q^2)|^2$  and a minimum at $\cos\theta=0$, which equals to $\tau |G_M(q^2)|^2$.
The relative contribution of the electric $\sigma_E$ (dashed lines) and magnetic $\sigma_M$ (solid lines) terms to the differential cross section
are illustrated in Fig. \ref{Fig:ang}a, for two values of $q^2=5$ and 8 GeV$^2$.

The total cross section is a quadratic combination of FFs, which does not contain interference terms:
\be
\sigma=\frac{\pi \alpha^2}{6m^2\tau\sqrt{\tau(\tau -1)}}\left( 2\tau |G_M|^2+|G_E|^2 \right ).
\label{eq:eqstot}
\ee
Let us introduce the angular asymmetry, ${\cal A}$, which enhances the different angular behavior of the 
electric and magnetic terms with respect to $\theta=90^\circ$. One can express the angular dependence of the differential cross section in terms of the angular asymmetry ${\cal A}$ as:
\be
\displaystyle\frac{d\sigma}{d(\cos\theta)}=
\sigma_0\left [ 1+{\cal A} \cos^2\theta \right ],
\label{eq:eqsa}
\ee
where $\sigma_0$, the differential cross section at
$\theta=\pi/2$,  and ${\cal A}$ can be written as functions of the FFs as:
\be
\sigma_0=\frac{\alpha^2}{4q^2}\sqrt{\frac{\tau }
{\tau -1}}
\left (|G_M|^2+ \frac{1}{\tau}|G_E|^2\right );
~{\cal A}=\displaystyle\frac{\tau|G_M|^2-|G_E|^2}{\tau|G_M|^2+|G_E|^2}=
\displaystyle\frac{\tau-{\cal R}^2}{\tau+{\cal R}^2},~{\cal{R}} = \frac{|G_{E}|}{|G_{M}|}.
\label{eq:eqa}
\ee
The angular asymmetry ${\cal{A}}$ lies in the range $-1\le {\cal A}\le 1$. For $G_E=0$ one obtains  ${\cal A}=1$ and for $G_E=G_M$ one obtains ${\cal  A}=(\tau-1)/(\tau+1)$.

The electric and magnetic contributions to the total cross section and to the angular asymmetry are illustrated in Fig. \ref{Fig:ang}b, as function of $q^2$. The unphysical region is indicated by a dashed area. Although the magnetic contribution largely dominates in the physical region, the relative contribution of the electric term to the cross section is larger than 10\% for $q^2\le 15$ GeV$^2$, and it is even larger for the angular asymmetry.

From the total cross section, it is possible to extract $|G_M|$ under a definite hypothesis on the ratio.  The 
experimental results are usually given 
in terms of $|G_M|$, under the hypothesis that $G_E=0$ or $|G_E|=|G_M|$. The first hypothesis is 
arbitrary whereas the second one is strictly valid at threshold only, and there is no 
theoretical argument which justifies its validity at any other momentum 
transfer, where $q^2\neq 4m^{2}$. However, $G_E$ plays a minor role in the cross section and different hypothesis for  
$|G_E|$ do not affect strongly the extracted values of $G_M$, due to 
the kinematical factor 
$\tau$, which weights the magnetic contribution to the differential cross 
section and makes the contribution of the electric FF to the cross section smaller and smaller as $q^2$ increases. 

The individual determination of the FFs in time-like region has not yet been done. The ratio $R=G_E/G_M$ has been determined from a two parameter fit of the differential cross section, by PS170 at LEAR \cite{Ba94}, and more recently  by the BABAR Collaboration using initial state radiation, $e^++e^-\to \overline{p}+p+\gamma$ \cite{Babar}. Data are very limited and affected by large errors, mainly due to statistics. The results from BABAR suggest a ratio larger than one, in a wide region above threshold, whereas data from \cite{Ba94} suggest smaller values. The results from \PANDA are expected to clarify this issue.

\subsection{Discussion}
\begin{figure}
\begin{center}
\includegraphics[width=10cm]{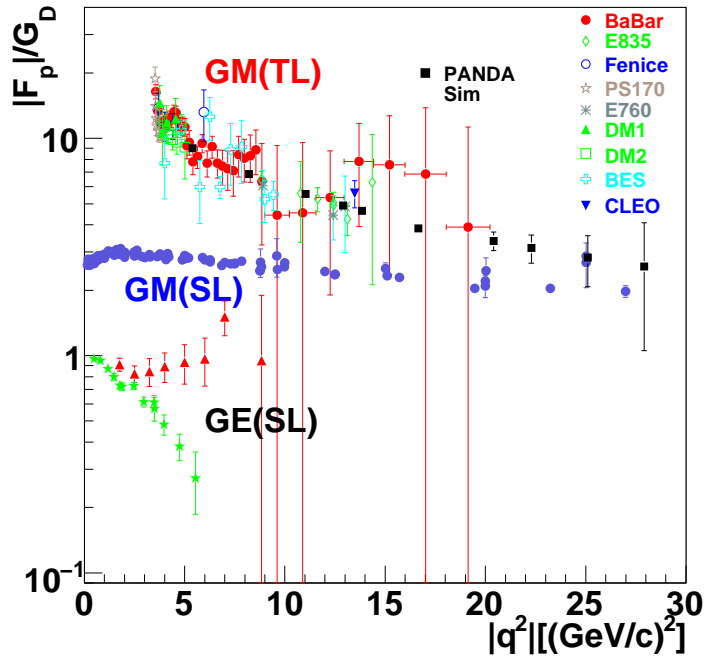}
\vspace*{.2 truecm}
\caption{World data on proton form factors, in time and space-like regions, as functions of $|q^2|$, rescaled by dipole. From top to bottom, magnetic FF in time-like region including \PANDA simulated results (black solid squares), magnetic FF in space-like region (blue circles), electric FF in space-like region, from unpolarized (red triangles) and polarized (green stars) experiments.}
 \label{Fig:TLSL}
 \end{center}
\end{figure}

A general illustration of the world data for the proton form factors is given in Fig. \ref{Fig:TLSL}, where proton FFs are shown as function of $|q^2|$, allowing a straightforward comparison in the whole kinematical region. In order to eliminate the steep $q^4$ dependence, all FFs are rescaled by the dipole function. 

From top to bottom, one can see the magnetic proton FF in TL region, under the assumption $|G_E|=|G_M|$, the magnetic proton FF in SL region which is obtained for $q^2\le 8.8 $ GeV$^2$ obtained under the assumption $|G_E|=0$ (blue circles) and the electric FF in SL region. Two series of data clearly show the discrepancy between unpolarized (red triangles) and polarized (green stars) measurements. 

The expected precision of the future measurements with \PANDA (black solid squares) is shown in comparison with the existing data. For \PANDA each point corresponds to an integrated luminosity of 2 fb$^{-1}$, which can be obtained in four months of data taking. These results have been obtained in frame of Montecarlo simulations, which takes into account the geometry of the detector, efficiency and acceptance and is based on a realistic parametrization of FFs  \cite{PANDA}. One can see that \PANDA will cover a large kinematical range and bring useful information with respect to the determination of the asymptotic region.

A more direct representation of FFs is given by the Dirac $F_1$ and Pauli $F_2$ FFs, which are linear combinations of $G_E$ and $G_M$. PQCD predicts the asymptotic behavior $F_1\sim Q^{-4}$ and $F_2\sim Q^{-6}$ which is followed by the Rosenbluth measurements, but not compatible with polarization data, which suggest instead the following ratio: $F_2/F_1\sim Q^{-1}$.

The values of $G_M$ in the TL region, 
obtained under the assumption  $|G_E|=|G_M|$, are larger 
than the corresponding SL values. A difference up to a factor of two in the absolute values in SL and TL regions can be seen also for other hadron FFs, including pions and neutrons, up to the largest value at which TL FFs have been measured. 
This has been considered as a proof of the non applicability of the Phr\`agmen-Lindel\"of theorem, 
or as an evidence that the asymptotic regime is not reached \cite{Bi93}. The Phragm\`en-Lindel\"of theorem constrains definitely FFs in 
TL and in SL regions to have the same value at large $q^2$. This theorem has other applications in particle physics, such as, for example, the Pomeranchuk theorem, concerning the asymptotic behavior of the total cross sections for $a+b$ and $\bar a +b$ collisions ($a$ and $b$ any hadrons): $\sigma_T(ab)=\sigma_T(\bar a b)$. However, to be rigorous, the applicability of this theorem to FFs, which seems evident, has not yet been proved. 

In principle, asymptotic properties should be discussed for $F_1$ and $F_2$.
The analyticity of FFs allows to apply the Phragm\`en-Lindel\"of 
theorem which gives a rigorous prescription for the asymptotic behavior of 
analytical functions: 
\begin{equation}
\lim_{q^2\to -\infty} F^{(SL)}(q^2) =\lim_{q^2\to \infty} F^{(TL)}(q^2).
\label{eq:eqph}
\end{equation}
This means that, asymptotically, FFs have the following constraints: 1) the imaginary part of FFs, in TL region, vanishes: $ Im F_i(q^2)\to 0,$ as $q^2\to \infty $; 2) the real part of FFs, in TL region, coincides with the 
corresponding value in SL region, because FFs are real functions in SL region, due to the hermiticity of the corresponding electromagnetic Hamiltonian.

Unfortunately, this theorem does not allow to indicate the physical value of $q^2$, starting from which it is working at some level of precision. For this aim one needs some additional dynamical information. The assumption of the analyticity of FFs allows to connect the nucleon FFs in SL and in TL regions and to extend a parametrization of FFs available in one kinematical region to the other kinematical region. Dispersion relation approaches, which are based essentially on the analytical properties of nucleon electromagnetic FFs, can be considered a powerful tool for the description of the $q^2$ behavior of FFs in the entire kinematical region. The vector meson dominance (VDM)  models, can be also extrapolated from the SL region to the TL region (see \cite{ETG05} and Refs. therein). The quark-gluon string model \cite{Ka00}  allowed firstly to find the $q^2$ dependence of the electromagnetic FFs in TL region, in a definite analytical form, which can be continued in the SL region.

In order to test these requirements, the knowledge of the differential cross section for $e^++e^-\leftrightarrow p+\bar{p}$ is not sufficient, and polarization phenomena have to be studied. In this respect, T-odd polarization observables, which are determined by $Im F_1F_2^*$, are especially interesting. The simplest of these observables is the $P_y$ component of the proton polarization in $e^++e^-\to p+\bar{p}$ that in general does not vanish, even in collisions of unpolarized leptons, or the asymmetry of leptons produced in $p+\bar{p}\to e^++e^-$, in the collision of unpolarized antiprotons with a transversally polarized proton target (or in the collision of transversally polarized antiprotons on an unpolarized proton target) \cite{Bi93,ETG05}. These observables are especially sensitive to the different parametrizations of FFs, and suggest that the corresponding asymptotics are very far \cite{ETG05a}.

\section{Two photon exchange}
\label{sec:twophoton}

As stressed in the introduction, the expressions of the cross section Eqs. (\ref{eq:sigma}, \ref{eq:eq2}) assume one-photon exchange. In principle, the interaction can occur through two (or more) photon exchange (TPE). Although if such mechanisms are suppressed by powers of $\alpha$ they could play a role  at large $q^2$, due to possible enhancement from a mechanism where the momentum is equally shared between the photons.
In such case the decrease of the cross section due to $\alpha$ counting would be compensated by the steep decrease of FFs with $q^2$. Recently, the possibility of a sizable TPE contribution has been discussed as possible solution to discrepancies between experimental data, on elastic electron deuteron scattering  \cite{Re99} and elastic electron proton scattering \cite{Pe07}. 

The model independent analysis of experimental observables taking into account the TPE contribution,  for $ep$ scattering and for the crossed annihilation channels can be found in Ref. \cite{Re04,Ga06,Ga05}. The presence of TPE induces a more complicated spin structure of the matrix amplitude. In the scattering channel, instead that two real FFs, functions of one kinematical variable, $q^2$, one has to determine three FFs, complex functions of two kinematical variables, and the $\epsilon$ linearity of the Rosenbluth formula does not hold anymore. However, it is still possible to measure the real FFs, using electron and positron scattering on proton, in the same kinematical conditions, or measuring three T-odd or five T-even polarization observables. In the annihilation channel TPE induces four new terms in the angular distribution, of the order of $\alpha $ compared to the dominant contribution and which are odd in $\cos\theta$. 

Therefore, the non linearity of the Rosenbluth fit in the scattering channel d the presence of odd $\cos\theta$ terms in the annihilation channel can be considered as a model independent signatures of TPE (more exactly, of the real part of the interference between one and two photon exchange). Evidence of TPE, based on these signatures has not been found in the experimental data on electron elastic scattering on particles with spin zero \cite{Ga08}, one half \cite{ETG}, and one \cite{Re99}. An analysis of the BABAR data \cite{Babar} also does not show the evidence of two photon contribution \cite{ETG08}. On the basis of simulation studies, it can be shown that the future \PANDA experiment will be sensitive to a TPE contribution $\ge 5\%$ of the main (one photon) contribution \cite{epjhal}.

Let us stress that the main advantage of the search of TPE in TL region is that the information is fully contained in the angular distribution. TPE effects cancel (are singled out) in the sum (difference) of the cross section at complementary angles, allowing to extract the moduli of the true FFs \cite{Ga06,Ga05}. TPE effects also cancel if one does not measure the charge of the outgoing lepton.

\section{Conclusion}

In the next future the knowledge of electromagnetic proton FFs will be extended in a wide kinematical region, allowing a unified description in both SL and TL regions. It will be possible to clarify both issues, the reaction mechanism and the proton electromagnetic structure at short distances. In particular, the individual determination of FFs will be possible in TL region, for moderate $q^2$ values. These data are expected to constrain nucleon models. For $q^2\ge 20$ GeV$^2$, where the electric contribution should become negligible, the validity of asymptotic properties predicted by QCD will be tested. Note that there is no principal reason for which the electric and the magnetic FFs should reach the same asymptotics, at the same $q^2$.

Due to crossing symmetry properties, the reaction mechanism should be the same in SL and TL regions, at similar values of the transferred momentum. If TPE is the reason of the discrepancy between the polarized and unpolarized FFs measurements in SL region, a contribution of $5\%$ is necessary to bring the data in agreement in the $|q^2|$ range between 1 and 6 (GeV/c)$^2$. Such level of contribution will be detectable in the \PANDA experiment. In Ref.  \cite{By07} the discrepancy has been attributed to the method of calculating radiative corrections. Radiative corrections are specific to each of these reactions, therefore a comparison of the data issued from the three channels, $ep$ scattering, $e^+e^-$ and $\bar pp$ annihilation, will shed light on the reaction mechanism. 

This work has been initiated within a long term collaboration with Prof. M. P. Rekalo and it is largely based on common work with Dr. G.I. Gakh. The members of the \PANDA Group at IPN Orsay are acknowledged for interesting discussions and remarks.

\end{document}